\documentclass{PoS}

\usepackage{amsfonts}
\usepackage{amsmath}
\usepackage{subfigure} 

\title{
  One loop matching factors for staggered bilinear operators
with improved glue
}

\ShortTitle{
  One loop matching  
  for staggered bilinears with improved glue
}

\author{\speaker{Jongjeong Kim}, \ \  Weonjong Lee\\
  Frontier Physics Research Division and Center for Theoretical Physics \\
  Department of Physics and Astronomy, 
  Seoul National University, Seoul, 151-747, South Korea \\
  E-mail: \email{rvanguard@phya.snu.ac.kr}, \ \  
          \email{wlee@snu.ac.kr}}

\author{Stephen R. Sharpe\\
  Physics Department, University of Washington, Seattle, WA 98195-1560 \\
  E-mail: \email{sharpe@phys.washington.edu}}

\abstract{ We present results for matching factors for 
  bilinear operators composed of HYP-smeared staggered fermions and 
  constructed using   HYP-smeared fat links. The matching factors are
  calculated perturbatively at one-loop order. 
  The new feature of our calculation compared to previous work on
  HYP-smeared staggered fermions is
  the use of the Symanzik-improved gluon propagator, which allows
  our results to be applied to our ongoing simulations based on
  configurations generated by the MILC collaboration.
  We address the issue of the relative efficiency of
  various improvement schemes in reducing one-loop
  corrections to the matching factors.  }

\FullConference{
  The XXVII International Symposium on Lattice Field Theory - LAT2009\\
  July 26-31 2009
  Peking University, Beijing, China
}

\begin{document}

\section{Introduction} 
Matching factors are needed to
convert matrix elements calculated on the lattice into 
physical observables in continuum renormalization schemes.
We focus here on matching factors
for the bilinear operators that we are using as 
part of an ongoing numerical study using MILC configurations.
The valence fermions of which these operators are composed 
are HYP-smeared~\cite{Hasenfratz:2001hp} improved staggered fermions,
which thus differ from the sea quarks, 
which use the ``asqtad'' staggered action~\cite{milc-01-1}.
We calculate the matching factors 
(which are, in general, matrices)
using perturbation theory at one-loop order.
What is new compared to previous calculations using HYP-smeared
staggered 
fermions~\cite{DS88,Patel:1992vu,Ishizuka:1993fs,Lee:1994xs,Lee:2002ui,Lee:2003sk}
is the use of the Symanzik-improved gluon action rather than
the Wilson plaquette action.
This generalization is necessary because the MILC configurations
use an improved gluon action.

This work is part of a larger project, whose ultimate aim is
to provide matching factors needed for the calculation of both
quark masses \cite{Bae:2008qe} and $B_K$
\cite{wlee-09-1,wlee-09-2,wlee-09-3,wlee-09-4}.
The results presented here allow us to study the relative
impact of improving quark and gluon actions.

\section{Actions and Bilinear Operators}
Since sea-quarks enter first at the two-loop level 
(through vacuum polarization diagrams), 
we do not give the details of the sea-quark action.

For the valence quarks, we use the HYP-smeared staggered fermion action.
This has the same form as the unimproved staggered fermion action,
\begin{equation} 
  S_f = 
  \frac{1}{2} \sum_{n, \mu} \eta_{\mu}(n) \bar{\chi}(n) 
  \Big( 
  V_{\mu}(n) \chi(n + \hat{\mu}) - 
  V^{\dagger}_{\mu}(n - \hat{\mu} ) \chi(n - \hat{\mu}) \Big)
  + m \sum_{n} \bar{\chi}(n)\chi(n) 
  \,, 
\end{equation}
[with $n \in \mathbb{Z}^4$ a lattice coordinate and $\eta_\mu(n) =
(-1)^{n_1 + n_2 + \cdots + n_{\mu-1}}$]
but with the original thin links $U_\mu$ replaced with HYP-smeared
links $V_\mu$. For the details of HYP-smearing, see
Ref.~\cite{Hasenfratz:2001hp}.
This action has a number of important properties:
\begin{enumerate}
\item It substantially reduces the breaking of taste 
symmetry~\cite{Bae:2006cj,Bae:2007ec,Bae:2008qe}; 
\item It significantly reduces one loop
corrections~\cite{Lee:2002ui};
\item It performs tadpole
improvement automatically~\cite{Lee:2002fj};
\item It reduces scaling
violations efficiently~\cite{wlee-06-1}.
\end{enumerate}
Given also its simplicity, it is an attractive choice for
an improved staggered action.

The HYP-smeared links $V_\mu$ can be expressed in terms of 
blocked gauge fields $B_\mu$:
\begin{equation} 
  V_\mu(x) = \exp \big( i B_\mu ( x \!+\!\hat{\mu}/2) \big) \,. 
\end{equation} 
The blocked gauge fields $B_\mu$ can be written as a power series
in the original gauge fields $A_\nu$,
\begin{equation} 
  B_\mu = \sum_{n=1}^{\infty} B_\mu^{(n)}\,, 
\end{equation} 
where $B_\mu^{(n)}$ contains all terms with $n$ powers of the gauge
fields $A_\nu$. In the present one-loop calculation, we need only
the linear term, $B^{(1)}_\mu$.
One might have expected the quadratic term $B^{(2)}_\mu$ to enter as well,
but its contribution vanishes due to the fact that
HYP-smearing includes $SU(3)$ projections~\cite{Patel:1992vu,Lee:2002fj}.

The linear term can be written in terms
of a blocking kernel $h_{\mu\nu}$, which it is convenient to
define in momentum space:
\begin{equation}
  B^{(1)}_\mu (k) = \sum_\nu h_{\mu\nu}(k) A_\nu(k) \,.
\end{equation} 
All information about the HYP-smearing, including the 
smearing parameters, is contained in $h_{\mu\nu}$. 
Following Ref.~\cite{Lee:2002ui}, we decompose the kernel
into diagonal and off-diagonal parts
\begin{equation} 
  h_{\mu\nu}(k) = \delta_{\mu\nu} D_\mu(k) + (1 -
  \delta_{\mu\nu}) G_{\mu\nu}(k) \,, 
\end{equation} 
With smearing coefficients chosen to remove ${\cal O}(a^2)$ taste symmetry
breaking coupling at tree level
($\alpha_1=0.75$, $\alpha_2=0.6$ and $\alpha_3=0.3$
in the notation of Ref.~\cite{Hasenfratz:2001hp}),
the diagonal part is
\begin{equation} 
  D_\mu(k) = 1 -
  \sum_{\nu\ne\mu} {\bar s}_\nu^2 + \sum_{\nu < \rho \atop
    \nu,\rho\ne\mu}{\bar s}_\nu^2 {\bar s}_\rho^2 - {\bar s}_\nu^2
  {\bar s}_\rho^2 {\bar s}_\sigma^2 \,,
\end{equation} 
with $\bar{s}_\mu = \sin(k_\mu / 2)$,
while the off-diagonal part is 
\begin{equation} 
  G_{\mu\nu}(k) = {\bar s}_\mu
  {\bar s}_\nu \left[ 1 - \frac{({\bar s}_\rho^2 
      + {\bar s}_\sigma^2)}{2} 
    + \frac{{\bar s}_\rho^2 {\bar s}_\sigma^2}{3}
  \right] \,.
\end{equation} 
For unimproved staggered fermions the blocking kernel
simply reduces to $h_{\mu\nu}=\delta_{\mu\nu}$.
We now turn to the Symanzik-improved gluon action, which can be
written as
\begin{equation} 
  S_g = \frac{2}{g_0^2} 
  \bigg[ c_{\rm pl} \sum_{\rm pl} {\rm ReTr} (1 - U_{\rm pl}) 
  + c_{\rm rt} \sum_{\rm rt} {\rm ReTr} (1 - U_{\rm rt}) 
  + c_{\rm pg} \sum_{\rm pg} {\rm ReTr} (1 - U_{\rm pg}) 
  \bigg] \,. 
  \label{eq:sg} 
\end{equation} 
Here, pl, rt, and pg represent plaquette, rectangle and
parallelogram, respectively. The coefficients $c_i$, where $i$=pl,
rt, or pg, should be chosen so as to improve the
scaling behavior. In the present calculation, we use the 
tree-level improved coefficients
\begin{equation} 
  c_{\rm pl} = \frac{5}{3},
  \quad 
  c_{\rm rt} = -\frac{1}{12}, 
  \quad
  {\rm and}
  \quad c_{\rm pg} = 0 \,,
\end{equation} 
which were determined by L\"uscher and 
Weisz~\cite{Weisz:1982zw,Luscher:1984xn}. 
The MILC collaboration actually use the 1-loop improved values for the
coefficients, but this leads to changes which would enter
only into a two-loop calculation.
Note that the Wilson gauge action corresponds to
$c_{\rm pl} = 1$, $c_{\rm rt} = c_{\rm pg} = 0$.
For the lattice bilinears, we use the operators
of Ref.~\cite{KlubergStern:1983dg}, 
which reside on $2^4$ hypercube.
%
The operator with spin $S$ and taste $F$ can be written as
\begin{equation} 
  [ S \times F ](y) = \frac{1}{16}
  \sum_{A,B} [\bar{\chi}_b(y+A) 
  \ (\overline{ \gamma_S \otimes \xi_F} )_{AB} \ \chi_c(y+B)] 
  \ {\cal V}^{bc}(y+A,y+B) \,, 
\end{equation}
where $y$ is the coordinate of hypercube, and $A$ and $B$ denote the
corners of hypercubes. Gauge invariance is maintained
by the inclusion of ${\cal V}^{bc}(y+A,y+B)$, which is
constructed by averaging over the shortest paths connecting
$y+A$ and $y+B$, with each path constructed from the products of 
HYP-smeared links $V_\mu$. In this way the operators are improved
in the same fashion as the action. This also ensures that
the currents $[V \times S]$ and $[A \times P]$ are conserved.

\section{Improved Gluon Propagator}
Propagators and vertices necessary to calculate
perturbative corrections to the staggered bilinear operators can be
found in Refs.~\cite{DS88, Patel:1992vu, Ishizuka:1993fs,Lee:1994xs}, 
with the exception of the gluon propagator 
for the Symanzik-improved action. Thus we discuss only the latter here.

The improved gluon propagator was worked out originally in
Ref.~\cite{Weisz:1982zw}. We have found a convenient repackaging
of the result, which we present here. This uses the notation\footnote{%
Be careful that $\hat{k}^4 \neq (\hat{k}^2)^2$ in this notation.
}
\begin{equation}
  \hat{k}^n \equiv \sum_\mu \hat{k}_\mu^n \,,
  \qquad
  \hat{k}_\mu \equiv 2 \sin (k_\mu / 2) \,,
\end{equation}
and the following orthogonal projectors:
\begin{equation}
  {\cal P}_{\mu\nu} =
  \frac{\hat{k}_\mu \hat{k}_\nu}{\hat{k}^2}\,,
  \qquad
  \delta^T_{\mu\nu} =
  \delta_{\mu\nu} - {\cal P}_{\mu\nu}\,,
  \label{eq:pd}
\end{equation}
The inverse gluon propagator with  covariant 
gauge fixing can then be written
\begin{eqnarray}
  {\cal D}^{-1}_{\mu\nu} &=& 
  \frac{1}{\alpha} \hat{k}^2 {\cal P}_{\mu\nu}
  + f \hat{k}^2 \delta^T_{\mu\nu}
  - c {\cal M}_{\mu\nu} \,,
\\
  {\cal M}_{\mu\nu} &=& 
  \delta_{\mu\nu} \hat{k}_\mu^2 \hat{k}^2 
  - \hat{k}_\mu^3 \hat{k}_\nu 
  - \hat{k}_\mu \hat{k}_\nu^3
  + \frac{ \hat{k}_\mu \hat{k}_\nu \hat{k}^4 }
  { \hat{k}^2 } \,,
\end{eqnarray}
where $\alpha$ is the gauge-fixing parameter,
\begin{equation}
  f = (\omega - c' \hat{k}^2 - c \hat{k}^4 / \hat{k}^2) \,,
  \label{eq:f}
\end{equation}
and
\begin{equation}
  \omega = c_{\rm pl} + 8 c_{\rm rt} + 8 c_{\rm pg} \,,
  \qquad
  c = c_{\rm rt} - c_{\rm pg} \,,
  \qquad
  c' = c_{\rm pg} \,.  
\end{equation}
Since we use gauge-invariant operators the matching factors
are independent of $\alpha$ and we chose $\alpha=1$.
For the tree-level improved Symanzik action $\omega=1$, $c=-1/12$ and
$c'=0$.

Inverting ${\cal D}^{-1}_{\mu\nu}$, we find the improved gluon
propagator to be
\begin{equation}
  {\cal D}_{\mu\nu} =
  \alpha \frac{{\cal P}_{\mu\nu}}{\hat{k}^2}
  + \frac{
    \left[
      \hat{k}^2 (\hat{k}^2 - \tilde{c} x_1) +  \tilde{c}^2 x_2 
    \right]
    \delta^T_{\mu\nu}
    + \tilde{c} (\hat{k}^2 - \tilde{c} x_1) {\cal M}_{\mu\nu}
    + \tilde{c}^2 ({\cal M}^2)_{\mu\nu}
  }
  {
    f 
    \left\{ 
      \hat{k}^2
      \left[
        \hat{k}^2 (\hat{k}^2 - \tilde{c} x_1) +  \tilde{c}^2 x_2 
      \right]
      - \tilde{c}^3 x_3
    \right\}
  } \,,
\end{equation}
where $\tilde{c} = c / f$, and
\begin{subequations}
  \begin{align}
    x_1 &= 
    {\rm Tr}({\cal M}) 
    = (\hat{k}^2)^2 - \hat{k}^4 
    =  2 \sum_{\mu < \nu} \hat{k}_\mu^2 \hat{k}_\nu^2 \,,  
    \\
    x_2 &= 
    \frac{1}{2}\left[{\rm Tr}^2({\cal M}) - {\rm Tr}({\cal M}^2)\right]
    = \hat{k}^2
    \left[
      \hat{k}^6 - (3/2)\hat{k}^2\hat{k}^4 + (1/2)(\hat{k}^2)^3
    \right]
    =  3 \hat{k}^2 
    \sum_{\mu < \nu < \rho} 
    \hat{k}_\mu^2 \hat{k}_\nu^2 \hat{k}_\rho^2 \,, 
    \\
    x_3 &= 
    \frac{1}{6}\left[
      {\rm Tr}^3({\cal M}) 
      - 3 {\rm Tr}({\cal M}) {\rm Tr}({\cal M}^2)
      + 2 {\rm Tr}({\cal M}^3)
    \right]
    \nonumber \\ 
    &= \frac{(\hat{k}^2)^2}{6}
      \left[
        (\hat{k}^2)^4 
        + 3 (\hat{k}^4)^2
        - 6 \hat{k}^4 (\hat{k}^2)^2
        + 8 \hat{k}^6 \hat{k}^2
        - 6 \hat{k}^8 
      \right]
    =  4 (\hat{k}^2)^2 
    \hat{k}_1^2 \hat{k}_2^2 \hat{k}_3^2 \hat{k}_4^2 \,.
  \end{align}
\end{subequations}

\section{Renormalization of Bilinear Operators}

\begin{figure}[htbp!] \centering
  \subfigure[]{\includegraphics[width=0.25\textwidth]{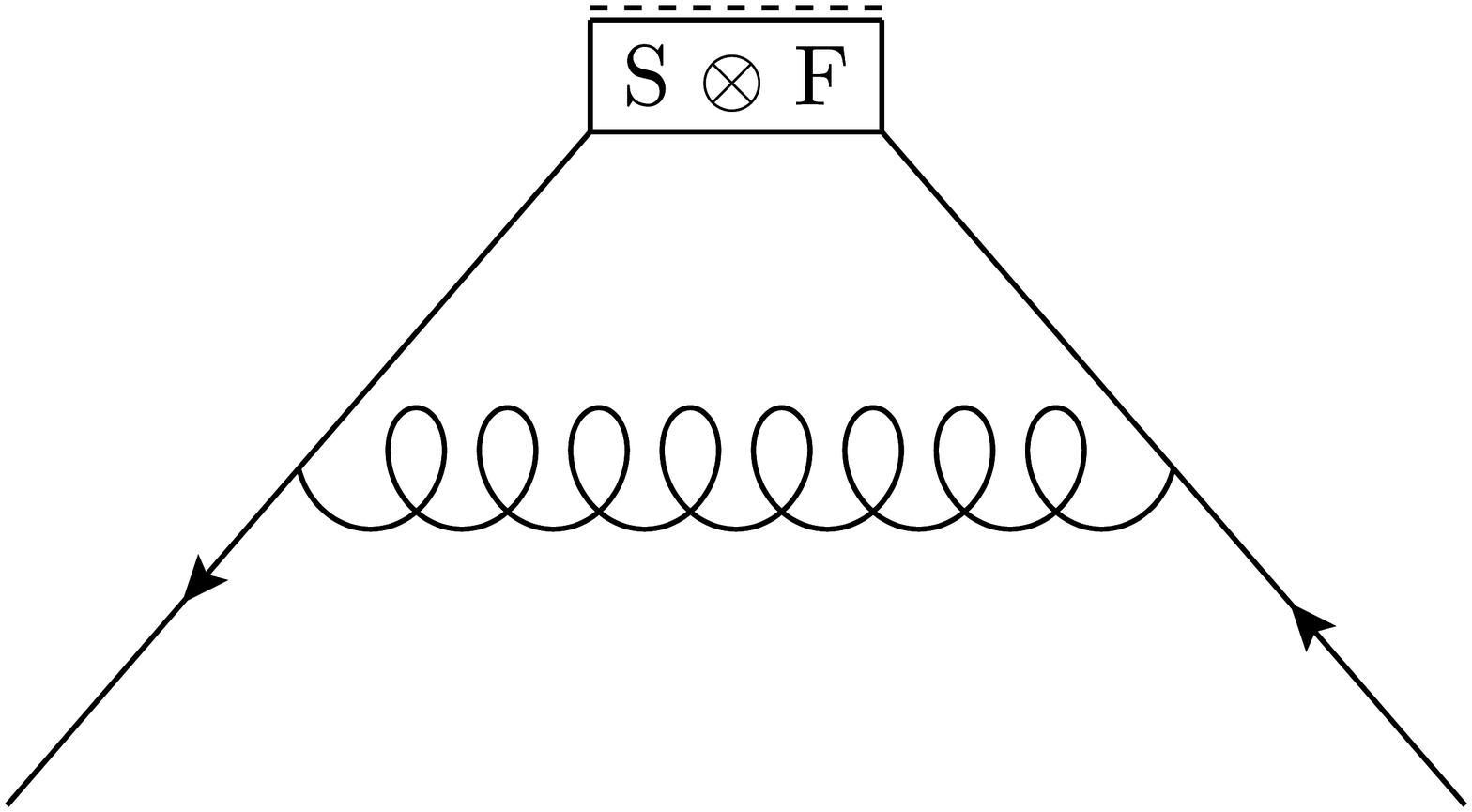}}
  \subfigure[]{\includegraphics[width=0.25\textwidth]{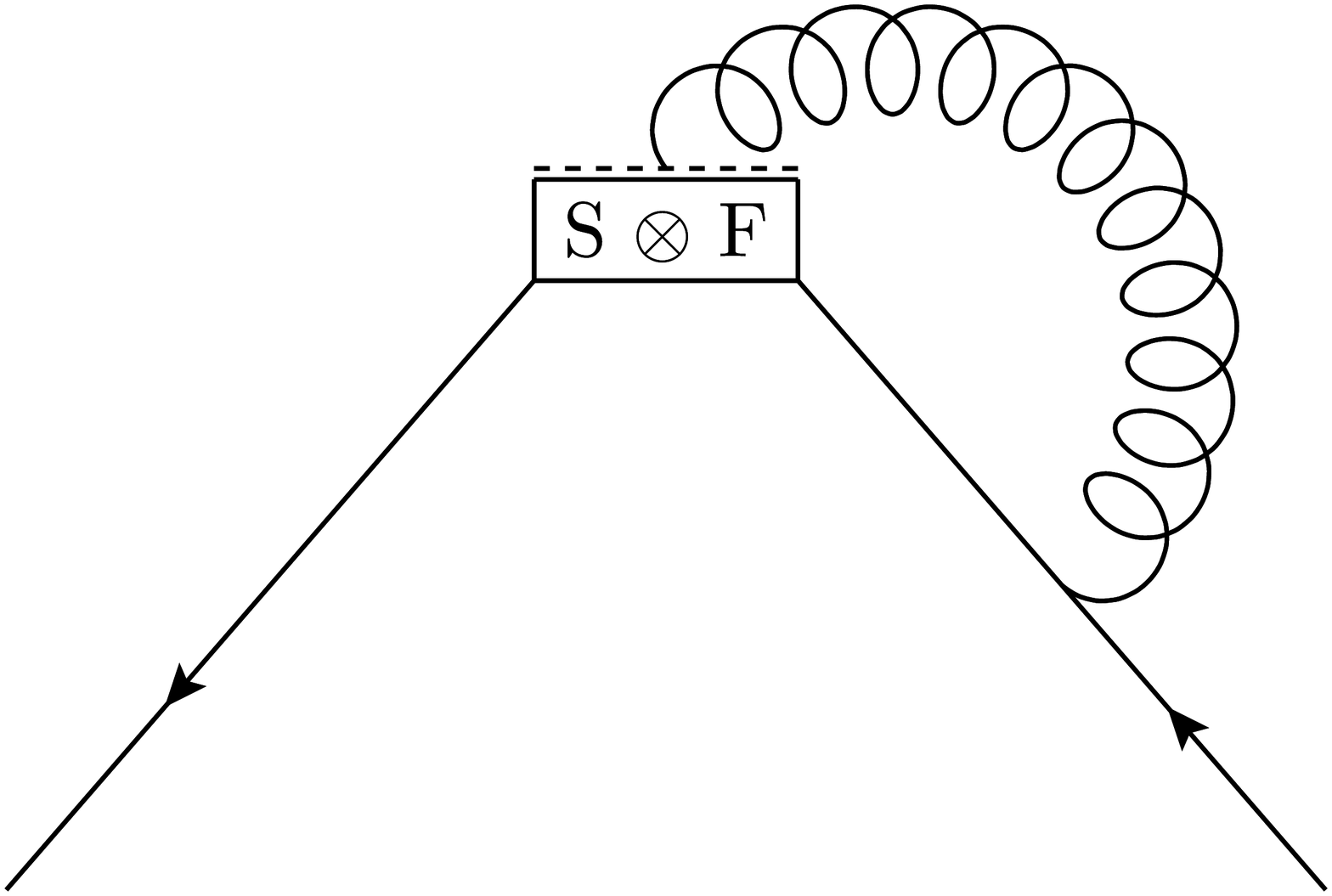}}
  \subfigure[]{\includegraphics[width=0.25\textwidth]{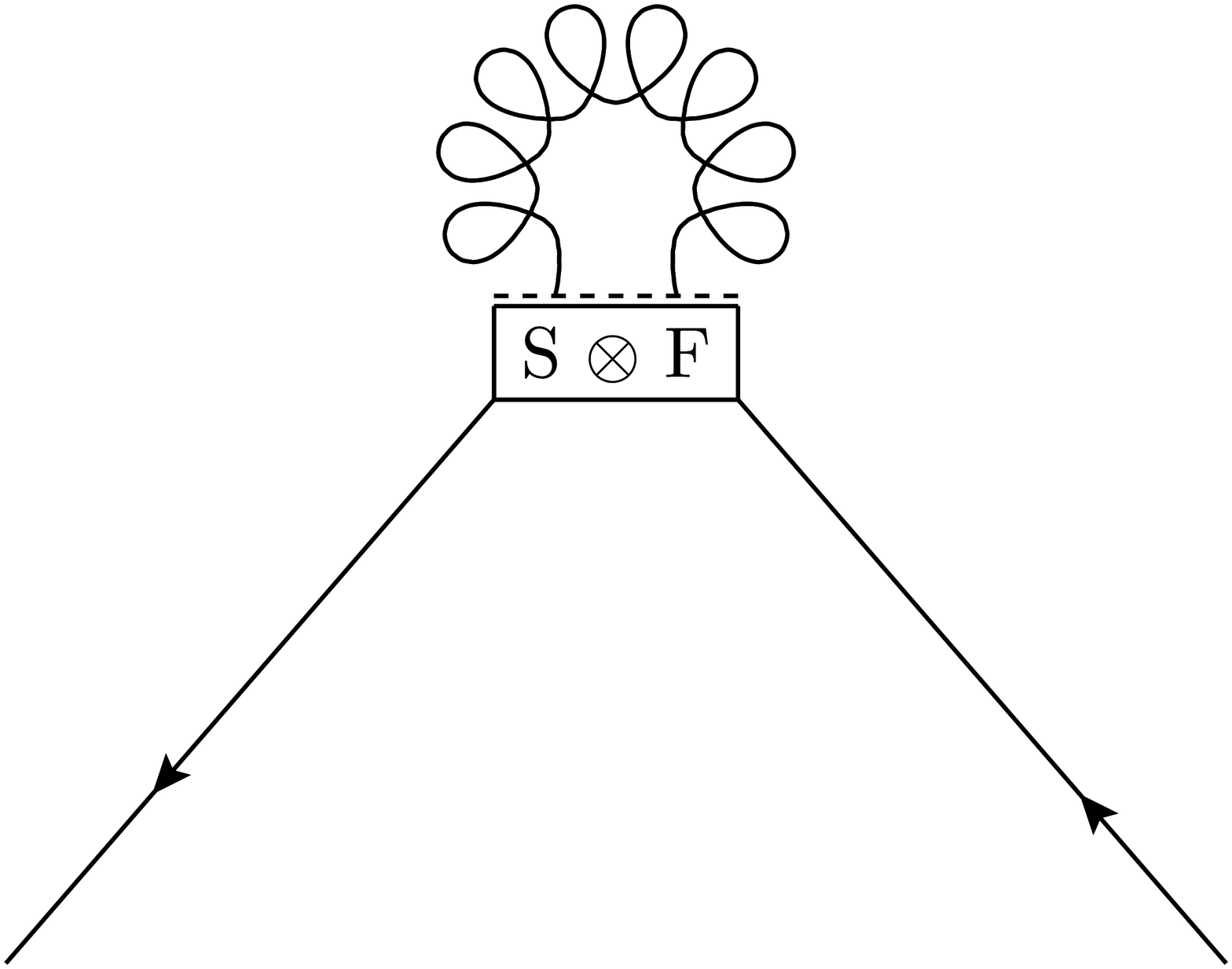}}
  \subfigure[]{\includegraphics[width=0.25\textwidth]{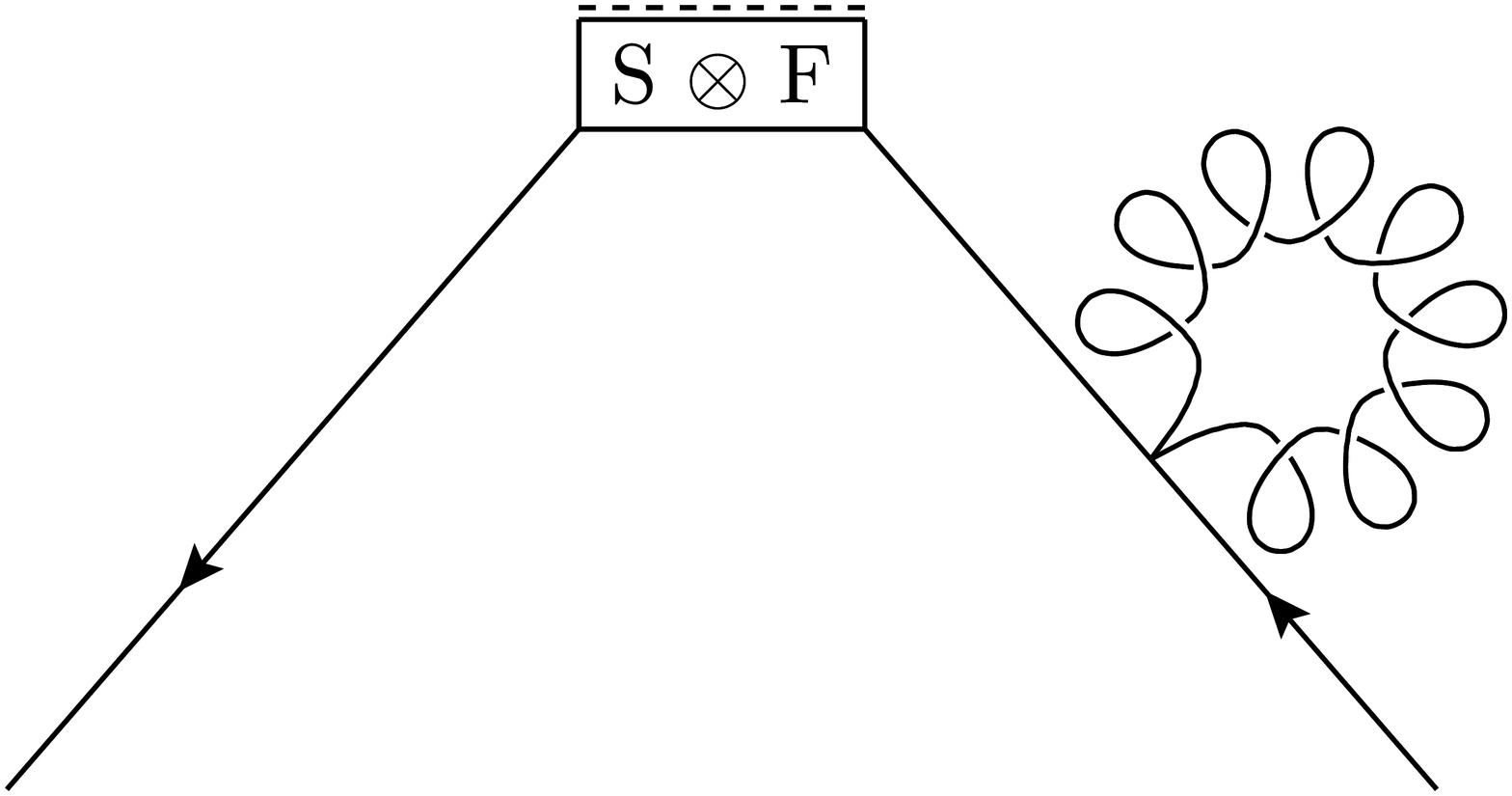}}
  \subfigure[]{\includegraphics[width=0.25\textwidth]{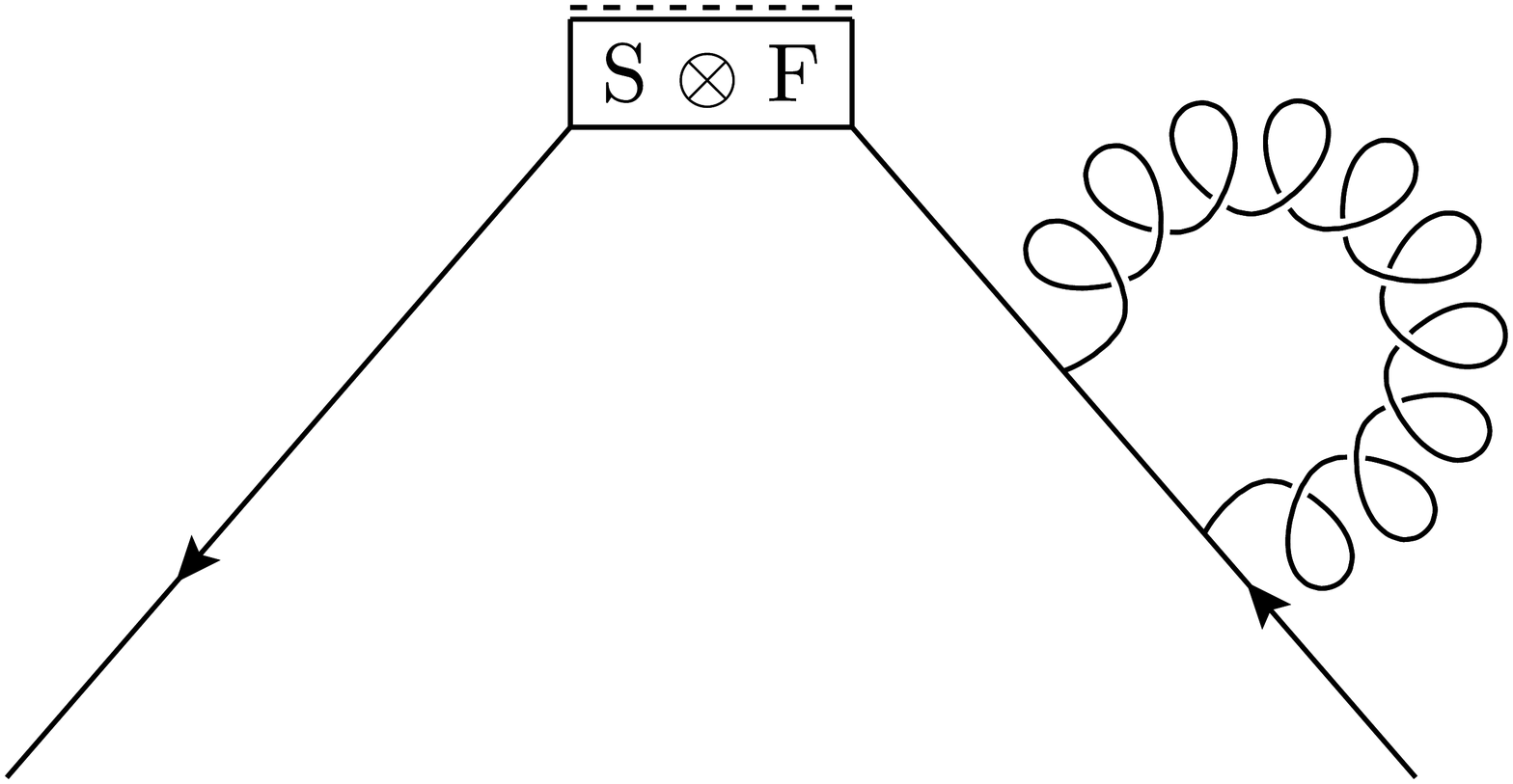}}
  \caption{Feynman diagrams for bilinear operators \label{fig:bi-op}}
\end{figure}

The Feynman diagrams relevant to the one-loop renormalization of the
bilinear operators are shown in Fig.~\ref{fig:bi-op}. Analytic
expressions for these diagrams using the unimproved gluon
propagator can be found in Refs.~\cite{Lee:2002ui,Lee:2003sk}.
It turns out that the generalization needed when
using the improved gluon propagator is relatively simple.
One simply replaces the composite
gluon propagator from a smeared link in the $\mu$'th
direction to a smeared  link in the $\nu$'th direction
(which is a building block of the calculation)
as follows:
\begin{equation} 
  (1/\hat{k}^2)\sum_\lambda h_{\mu\lambda}h_{\nu\lambda} \to
  \sum_{\alpha\beta} h_{\mu\alpha}h_{\nu\beta} {\cal D}_{\alpha\beta} 
\end{equation}
The key simplification arises from the fact
that this composite propagator contains off-diagonal terms {\em even
when the gluon propagator is diagonal (as it is for
the Wilson gauge action in Feynman gauge)}, 
so that no new types of contribution arise when moving to
the (off-diagonal) improved gluons propagator. 
One must simply evaluate the loop integrals 
(which is done numerically) using the
more complicated composite gluon propagator.

Two independent calculations have been done as a cross-check.
In the end, we obtain the one-loop renormalization factors of the
bilinear operators,
\begin{equation}
  {O}^{\text{Latt},(1)}_i =
  \left\{\delta_{ij} + \frac{4}{3}\frac{g^2}{(4\pi)^2} \Big[
    \gamma_{ij} \log( a \lambda ) +
    C^{Latt}_{ij}
    \Big]\right\}
  {O}^{\text{Latt},(0)}_j + {\cal O}(a) \,,
  \label{eq:ren}
\end{equation}
where the superscripts indicate the order in perturbation theory,
$i$ and $j$ are indices which run over the 256 bilinear operators,
$\lambda$ is the ``gluon mass'' used to regularize the infrared divergences,
and $\gamma_{ij}$ is the anomalous dimension matrix. 
The quantities of interest are
$C^{Latt}_{ij}$, the finite parts of the corrections.
These have, in general, non-zero off-diagonal elements.
A similar expression holds in continuum regularizations, and
by combining with the result (\ref{eq:ren}) one can obtain
the desired one-loop matching factors.
Full details will be presented in Ref.~\cite{our_paper}.
Here we simply quote,
in Tables \ref{tab:cii} and \ref{tab:cij}, 
results for some representative diagonal and off-diagonal
elements of $C^{Latt}_{ij}$. 

As noted in Ref.~\cite{Lee:2002ui}, one can apply 
the mean-field improvement method of Ref.~\cite{Lepage:1992xa}
to our HYP-smeared operators. This should reduce the size of
fluctuations in the gauge links. This
results in shifts in the values of the diagonal
elements $C^{Latt}_{ii}$~\cite{Patel:1992vu,Ishizuka:1993fs,Lee:1994xs}.
Mean-field improved values of $C^{Latt}_{ii}$ are shown in
the last four columns of Table \ref{tab:cii}.
\begin{table}[ht]
  \begin{center}
    \begin{tabular}{l|rrrr|rrrr}
      \hline
      \hline
      Operator  & $(a)$ \qquad &$(b)$ \qquad &$(c)$ \qquad &$(d)$ \qquad& 
      $(a)'$ \qquad &$(b)'$ \qquad &$(c)'$ \qquad &$(d)'$ \qquad \\ 
      \hline 
  $(1 \otimes 1)$                             
  &  41.73  &  2.60 &  32.82 &  1.91
  &  31.86  &  1.54 &  25.59 &  1.18 \\
  $(\gamma_5 \otimes 1)$                             
  &  -35.86  & -7.17 & -27.24 & -6.01
  &  -6.25  & -4.01 & -5.55  & -3.84 \\  
  $(\gamma_\mu \otimes 1)$                    
  &   0.00  &  0.00 & 0.00    & 0.00 
  &   0.00  &  0.00 & 0.00    & 0.00 \\
  $(\gamma_\mu \otimes \xi_{5})$              
  &  -22.51 & -3.97 & -17.04 & -3.10
  &  -2.77  & -1.86 & -2.58  & -1.65 \\
  $(\gamma_{\mu\nu} \otimes 1)$               
  &  -10.97 & -1.84 & -8.38  & -1.34
  &   -1.10 & -0.79 & -1.15  & -0.62 \\
  $(\gamma_{\mu\nu} \otimes \xi_{\rho\sigma})$
  &  -34.05 & -5.19 & -25.64 & -3.82  
  &  -4.44  & -2.03 & -3.95 & -1.65 \\ 
  \hline
  \hline
\end{tabular}
\end{center}
\caption[]{ 
    Diagonal part of representative diagonal coefficients 
    $C_{ii}^\text{Latt}$.
    Note that $\mu$, $\nu$, $\rho$, and $\sigma$ are all
    different. Results are given for the following four choices of
    actions: $(a)$ thin links (in the fermion action and bilinears) 
    with the Wilson plaquette action; $(b)$
    HYP-smeared links with the Wilson plaquette action; $(c)$ thin links with
    the improved gluon action; $(d)$ HYP-smeared links with the improved
    gluon action. The prime in the labels indicates that
    mean-field improvement is applied. Results are accurate
    to the number of digits quoted.
  }
  \label{tab:cii}
\end{table}
  
\begin{table}[ht]
  \begin{center}
    \begin{tabular}{cllrrrr}
  \hline\hline
  Name &  Operator-$i$  & Operator-$j$ &
   $(a)$ \quad &$(b)$ \quad &$(c)$ \quad &$(d)$ \quad \\
  \hline 
  $c_{VVM}$ & 
  $(\gamma_\mu \otimes \xi_\nu)$ & 
  $(\gamma_\mu \otimes \xi_\mu)$ &
  -3.042 & -0.351 & -2.495 & -0.321 \\
  $c_{VAM}$ & 
  $(\gamma_\mu \otimes \xi_{\mu5})$ & 
  $(\gamma_\mu \otimes \xi_{\nu5})$ &
  0.647 & 0.257 & 0.609 & 0.244 \\
  $c_{VTM}$ & 
  $(\gamma_\mu \otimes \xi_{\mu\nu5})$ &
  $(\gamma_\mu \otimes \xi_{\rho\nu5})$ &
  1.486 & 0.280 & 1.292 & 0.266 \\
  $c_{TAM}$ &  
  $(\gamma_{\mu\nu} \otimes \xi_{\mu5})$ &
  $(\gamma_{\mu\nu} \otimes \xi_{\rho5})$ &
  0.676 & -0.006 & 0.547 &  -0.003 \\
  \hline\hline
\end{tabular}
\end{center}
\caption[]{
  Off-diagonal coefficients $C_{ij}^{lat}$.
  The notation is the same as in Table
  \ref{tab:cii}. Results are accurate to the number of
  digits quoted.
}
  \label{tab:cij}
\end{table}

\section{Discussion}

We can use the results to compare the reduction in the
size of one-loop matching factors achieved by different
improvement schemes. For vector currents and the
off-diagonal coefficients, the anomalous dimensions vanish
and the $C_{ij}^{lat}$ give a direct measure of the size of the
corrections. Looking at the results for $(\gamma_\mu\otimes \xi_5)$,
a vector current containing 3 links, one sees that the corrections
in column (a) for unimproved fermions and glue are reduced by all choices 
of improvement scheme.
Comparing the impact of applying each improvement alone, 
the greatest reduction is achieved by mean-field
improvement, with HYP-smearing following close behind, 
but with the improved gauge action alone leading to 
a much smaller reduction.
Combining all three improvements leads to the smallest coefficients.
Similar results holds for the off-diagonal coefficients of
Table~\ref{tab:cij}, except that these are unaffected by mean-field
improvement.

For the scalar and tensor operators, which have non-zero anomalous
dimensions, the coefficients $C_{ij}^{lat}$ depend on the choice
of infrared regularization. Thus one should consider the difference between
the coefficients for fixed spin and differing tastes, i.e.
the differences between the first and second rows and between
the fifth and sixth rows of Table \ref{tab:cii}. 
For these differences one finds
a similar pattern of improvements to those noted above, except that for
the scalar bilinears HYP smearing leads to a significantly
greater reduction than mean-field improvement.

Overall, we conclude that improving the gluon action
reduces the one-loop corrections somewhat but is not nearly
as effective in this regard as HYP-smearing.

\section{Acknowledgments}
The research of J.~Kim and W.~Lee is supported by the Creative Research
Initiatives program (3348-20090015) of the KOSEF grant funded by the
Korean government (MEST). 
The work of S.~Sharpe is supported in part by the US DOE grant
no.~DE-FG02-96ER40956.
%

\end{document}